\documentclass{article}

\usepackage{graphicx}
\usepackage{psfig}
\usepackage{epsfig}
\usepackage[round]{natbib}

\title{Weather forecasts, Weather derivatives, Black-Scholes, Feynmann-Kac and Fokker-Planck}

\author{Stephen Jewson\footnote{\emph{Correspondence address}: RMS, 10 Eastcheap,
London, EC3M 1AJ, UK. Email: \texttt{x@stephenjewson.com}}\\
RMS, London, United Kingdom}
\begin{document}

\newcommand{\bx}[1]{\fbox{\begin{minipage}{15.8cm}#1\end{minipage}}}

\maketitle

\begin{abstract}
We investigate the relationships between weather forecasting,
weather derivatives, the Black-Scholes equation, Feynmann-Kac theory and the Fokker-Planck equation. 
There is one useful result, but
on the whole the relations we present seem to be 
more \emph{interesting} than \emph{practically useful}.

\end{abstract}

\section{Introduction}

Weather forecasting is all about predicting what the weather is
going to do next, or over the next 10 days. Weather derivatives
are a way of insuring oneself against adverse weather conditions.
The Black-Scholes equation is a financial model for the pricing of
certain kinds of financial contracts.
Feynmann-Kac theory
is an esoteric result from the study of partial differential 
equations and stochastic calculus. 
The Fokker-Planck equation is a result from statistical physics
concerning the probability density of the location of a particle.
We
describe, in an informal way, how these five topics are related.

\section{Temperature forecasts as Brownian motion}
\label{tisbm}

Today is Monday, and we start by considering a weather forecast
that attempts to predict the average temperature for next Saturday
i.e.\ a 5 day forecast. To be specific, we will define average
temperature, as is convention, to be the midpoint of the minimum
and maximum temperatures during the 24 hour period of Saturday. We
will consider a single-valued forecast, rather than a forecast of
the whole distribution of possible temperatures. How should such a
forecast be interpreted in statistical terms? The most obvious
interpretation is that the forecast represents the \emph{expectation}, or
\emph{mean}, of the distribution of possible temperatures.

Tomorrow we will have a new forecast for Saturday's temperature,
which will then be a 4 day forecast. The new forecast will most
likely have a different value from today's forecast, and will probably be more accurate,
at least on average over many trials, since it is predicting less
far into the future. Come Wednesday we will have a new forecast
again, and so on, until the final forecast is delivered on
Saturday, and the final measurements become available at the end
of the day Saturday and we can look back and judge how well our
forecasts have performed.

We have considered a situation in which we get new forecasts once
a day. One could also imagine getting new forecasts more
frequently, and could move towards a limit in which new forecast
information is arriving continuously. Some of our mathematics will
use this limit to simplify the notation, but our intuition will be
based mostly on the idea of daily updates.

How do the forecasts we are considering change from day to day? It seems natural that
they could go either up or down. Furthermore, if we were sure, on Tuesday, that
the Wednesday forecast was going to be higher, then we should
really incorporate that information into our Tuesday forecast.
Having incorporated all the information we have into the Tuesday
forecast, it would seem likely that the Wednesday forecast could
be either higher or lower, but would be the same on average.
This suggests we could model the changes in the forecast as a
random change with mean zero. In statistics such a stochastic
process that doesn't go up or down on average, but jumps randomly,
is known as a \emph{martingale}. We offer three explanations for why we
think that it is reasonable to model weather forecasts as
martingales:
\begin{itemize}

    \item The intuitive explanation given above: if they were
    \emph{not} martingales then they would have a predictable component,
    which contains information that could have been included in
    the current forecast. We assume that the forecasts are
    produced by a rational and efficient process, and hence that
    this information \emph{is} included in the forecast and so it
    \emph{is} a martingale. We have previously called this assumption
    the 'efficient forecast hypothesis', as a parody on the
    'efficient market hypothesis', which is a similar assumption
    used to justify the modelling of jumps in share prices as
    being random.

    \item A more general intuitive explanation that any
    expectation, of anything, can only change randomly, otherwise
    it cannot be an expectation.

    \item A mathematical proof known variously as the 
    'tower law'~\citep{baxterr} or the 
    'law of iterated expectations'(\citet{bjork98}, page 33) 
    that shows that \emph{all} expectations are martingales.

\end{itemize}

Temperature is very often reasonably close to normally
distributed, and so are temperature forecasts. If temperature
forecasts are normally distributed, then \emph{changes} in temperature
forecasts are also normally distributed. 
This makes our model for temperature forecasts more specific: they
are \emph{gaussian} martingales.
In fact, there is only one gaussian martingale and that is
Brownian motion, sometimes known as a Wiener process or a continuous random walk. 
So now we have arrived at the conclusion that
weather forecasts can be modelled using Brownian
motion. We will write this as the stochastic differential equation (SDE):

\begin{equation}\label{bm}
  d\mu=\sigma dB
\end{equation}

where $\mu$ is the weather forecast, $B$ is Brownian motion and $\sigma$ is a volatility.

If we knew $\sigma$ we could integrate this equation forward in time as an ensemble
to give possible future values for forecasts between now and Saturday, 
and also the possible future values for temperature on Saturday.

So the next question is: what is $\sigma$, and how does it vary in time?

\section{Weather forecast volatility}\label{section3}

We start by considering the volatility between Monday and Tuesday. Our \emph{first}
model for this 5-day forecast to 4-day forecast volatility is that it is constant at
all times of year. 
However, a cursory
investigation of forecasts shows that they show a larger variance
in winter, corresponding to the larger variance of temperatures at that time. 
As a result of this, the \emph{changes} in forecasts from day to
day show a larger variance, and so the volatility is larger. 
This is an entirely predictable effect. Our
\emph{second} model for the volatility is therefore that the volatility is
deterministic, with a seasonal cycle corresponding to the
seasonal cycle of variance of observed temperatures.
We have used this model previously in~\citet{jewson03j}.

We will now question our second model of volatility. Is the
volatility really completely predictable, as we have assumed above, or does it change from day to
day? (to be clear, at this point we mean: does the 5 day forecast to 4 day forecast
volatility change as the starting and target day of the forecast move in time...we address
the question of whether the 5 day forecast to 4 day forecast volatility differs
from the 4 day forecast to 3 day forecast volatility later).

Are not some weather situations more predictable than others,
and doesn't that lead to periods of lower volatility in the forecast? It has been
shown that weather forecasts can predict, with some skill,
that the variance of the conditional distribution of temperature
varies with time (for example, see~\citet{palmert88}). 
Another way of saying this is that
the variance of forecast errors varies in time.
We have extended this to show that it is
possible to predict the volatility with skill in advance~\citep{jewsonz02a}. So from
this we see that the volatility shows days to day variations in
size which overlay the seasonal changes, and also that part of
these variations are predictable. 

So where does this leave us? We have a volatility that varies
seasonally, in a way that is predictable infinitely far in advance.
It also varies on shorter timescales in a way that could be
considered completely unpredictable from a long distance in
advance, but becomes partly predictable over short time scales
using weather forecast models. In
fact, however, our own research seems to show that the predictable short 
time-scale variations in forecast uncertainty
are rather small, and do not 
add much, if anything, to the skill of temperature forecasts
(see~\citet{jewsonbz03a} and~\citet{jewson03g}). 
As a result, short term changes in the volatility of the forecasts are
also small.
Because
they are small, and for the sake of being able to make some
progress with simple mathematics, we are going to ignore them.
This means we can write the volatility simply as a fixed function
of time.

We now consider variations in the size of the volatility between
Monday and Saturday i.e. the variation with lead time (now we are
comparing the 5 day forecast to 4 day forecast volatility with the
4 day forecast to 3 day forecast volatility). Are the
sizes of the changes in forecasts between Monday and Tuesday any
different from the sizes of changes between Friday and Saturday?
In other words, does the volatility change as we approach the target day?
It is interesting to pose this question to a few meteorologists: we have
done so, and found no consistency at all between the answers that
we received. Our (albeit brief) data analysis,
however, seems to show, rather interestingly, that these
volatilities are roughly constant with lead time~\citep{jewson02d}. 
Maybe there is
some underlying reason for this, we don't know. 
Either way, it certainly simplifies the modelling.
Combined with the model of the
changes in the forecasts as being Brownian motion, this gives us a
linear variation in forecast mean square error, which has been
noted many times (and is also more or less unexplained).

To summarise this section: we now have a model that says 
that a temperature forecast for a single target day 
follows a Brownian motion with deterministic volatility
as we approach the target day.
The volatility varies with forecast day (because it depends on the
time of year) but does not vary with lead time.

Our arguments in this section have been somewhat ad-hoc: as 
an alternative
one could consider trying to develop models
for the volatility based on a detailed analysis of the
statistics of the variability of daily temperatures and daily
forecasts, and their dependencies. We have not tried this. 
It may be possible, but we suspect it will be very hard indeed, given how
hard it is to model just temperatures alone (see our attempts 
to solve this much simpler problem
in~\citet{caballerojb02} and~\citet{jewsonc03a}).

\section{Forecasts for the monthly mean}

We will now consider a slightly different weather forecast. It
is Monday 30th November. Tomorrow is the first day of December.
We are interested in a forecast for the average
temperature in December (defined, for the sake of being a little
precise, as the arithmetic mean of the daily averages). Today's
estimate of that average temperature is based on a single valued
forecast for the expected temperature that extends out for
the next 10 days, followed by estimates of the expected
temperature for the remainder of the month based on historical
data. Tomorrow (December 1st) we will update this forecast: we
will have one more day of forecasts relevant for the month of
December, and will need to use one fewer day of historical data.
The day after tomorrow we will get an observed value for December
1st (it won't be the final, quality controlled value, as so is still subject
to change, but we will
ignore that). 
We will then use 10 days of forecast from the 2nd to the 11th, and 
historical data from the 12th to the 31st.
The day after that we will get one more day of historical data, our forecast
will move forward one more step, and so on.
Then at some point the forecast starts to drop off the end,
and a few days after that we get the final observed data point for December
31st.

We can now think about how such a forecast changes in time.
In fact, the changes in the monthly
forecast are made up of sums of changes in daily forecasts, 
which we have already considered.
Since a sum of Brownian motions is a Brownian motion, we conclude
that we can model the \emph{monthly} forecast as a Brownian motion too.

The shape of the volatility for this monthly forecast Brownian motion, is, however,
a little more interesting than before. When the forecast first starts to
impinge upon December the volatility is going to be low. 
The forecast does not initially have much effect on the monthly mean, and so
the day to day changes in the monthly forecast are small. As more and more forecast takes
effect the volatility grows. During the bulk of the contract all
of the forecast is relevant, and the volatility has a relatively
constant value. Finally, at the end of the contract, the
volatility starts to reduce as less and less of the forecast is
used. What model could we use to represent this ramping up,
constant level, and then ramping down, of the volatility? The
actual shape of the ramps depends on the sizes of the volatilities
for daily forecasts, and of the correlations between changes of
the forecasts at different leads. To keep things simple we will
make two assumptions: firstly, that the sizes of the
volatilities for single day forecasts are constant...the same
assumption we made in section~\ref{section3}. 
This is partly justified by analysis of
data. Secondly, we assume that these changes are uncorrelated. This is not
really true...the changes in forecasts at different leads do show
some, albeit weak, correlations (see~\citet{jewson02d}). 
But it is not too far wrong, at least, and
allows us to make some progress. In particular, it means that the
ramping up and down is given by a straight line in volatility
squared. We will call this the 'trapezium' model. We have described this model
in detail in~\citet{jewson02a} and~\citet{jewson03j}, and used it in~\citet{jewson03k} and~\citet{jewson03l}.

\section{Statistical properties of weather forecasts}

We have concluded that both daily and monthly forecasts can be modelled using Brownian motion.
In the daily case the volatility is constant with lead, but varies seasonally. 
In the monthly case the volatility squared follows a trapezium shape. We can
now derive a number of further results from these models.
We start
with a consideration of \emph{transition probabilities}. 

We first ask the
question: if our forecast says $\mu(t)$ today (at time $t$) then what
is the distribution of the different things it might say at some point
in the future (at time $T$)? This can be solved by integrating equation~\ref{bm}
forward in time. This gives:

\begin{equation}
  \mu \sim N(\mu(t),\int_{t}^{T} \sigma^2 dt)
\end{equation}

i{.}e{.} the distribution of future values of our forecast is a normal distribution
with mean $\mu(t)$ and variance given by the integral of the daily volatilities squared.
Alternatively, the probabilities themselves as a function of the value
of the forecast $x$ and time $t$
are given by solutions of
the Fokker-Planck equation (also known as the Kolmogorov forward equation):

\begin{equation}
  \frac{\partial p}{\partial t}=\frac{\partial^2}{\partial x^2} (\sigma^2 p)
\end{equation}

along with a boundary condition based on what the forecast currently says.
This equation is a diffusion type of equation. The probabilities diffuse
outwards as time moves forwards.

Another related question is: if our forecast is going to say $x$ in the
future, what are the possible things it could be saying today?
These probabilities are given by the \emph{backward} Fokker-Planck equation
(also known as the Kolmogorov backward equation):

\begin{equation}
  \frac{\partial p}{\partial t}=-\sigma^2 \frac{\partial^2 p}{\partial x^2} 
\end{equation}

along with a boundary condition that specifies the final condition.
This equation is diffusion in reverse i{.}e{.} starting at the final 
forecasts, the probabilities diffuse outwards as time moves backwards.

For more details on these equations, see, 
for example, \citet{gardiner85} or~\citet{bjork98}.

\section{The Feynmann-Kac formula}\label{fk}

We now take a diversion from discussing weather forecasts, and
present the Feynmann-Kac formula. In subsequent sections we will
then see that this formula is related to the fair price of weather options.

Consider the following partial differential equation (PDE) for $F(x,t)$:

\begin{equation}\label{fkpde}
\frac{\partial F}{\partial t}+\mu\frac{\partial F}{\partial x}+\frac{1}{2}\sigma^2\frac{\partial^2 F}{\partial x^2}=0
\end{equation}

with the boundary condition:
\begin{equation}
F(x,T)=\Phi(x)
\end{equation}

This PDE is very "special": because of the particular nature of the terms in the equation 
we can solve it by taking expectations of a function of
Brownian motion. The Brownian Motion we must use is fixed by the coefficients 
in the PDE:

\begin{equation}
    dX=\mu dt+\sigma dB
\end{equation}

with the initial condition:
\begin{equation}
  X=x
\end{equation}

and the expectation we must calculate is
\begin{equation}
F(x,t)=E[\Phi(X_T)]
\end{equation}

What we see from this is that one interpretation of what the
solution of the PDE actually does is to somehow take expectations
of the final condition, at different points in time, and
according to some stochastic process. The proof of this relation
is easy, and is given in, for example, \citet{bjork98}.

\section{Weather derivative fair prices are martingales}
\label{fpismart}

We will now link our discussion of weather forecasts to the pricing of weather 
derivatives.
Weather derivatives are financial contracts that allow entities to
insure themselves against adverse weather. They are based on a
weather index, such as monthly mean temperature in December, and
have a payoff function which converts the index into a financial
amount that is to be paid from the seller of the contract to the
buyer of the contract. The payoff function can in principle have
any form, but a certain number of piecewise linear forms known as
swaps, puts and calls are most common (see, for example, \citet{jewson03a}).

The fair price of a weather derivative is defined, by convention,
as the expectation of the distribution of possible payoffs from
the contract. The fair price can form the basis of negotiations of
what the actual price should be. Typically the seller of the
contract (who sells such contracts as a business) would want to
charge more than the fair price, otherwise they could not expect
to make money on average. The buyer may be willing to pay above
the fair price because they are grateful for the chance to insure
themselves against potentially dire weather situations.

We can calculate the fair price using whatever relevant historical
data and forecasts are available to estimate the final distribution
of outcomes of the contract. 
This is the usual approach (details are given in~\citet{jewsonbz02a} and~\citet{brixjz02a}).
Alternatively, we can calculate the 
fair price by integrating equation~\ref{bm} foreward to the end of
the contract to give us the final distribution of settlement indices.

As we move through a contract, our estimate of the fair price will
change in time as the available historical data and forecasts change. The fair
price is an expectation, by definition, and we have argued in section~\ref{tisbm}
that expectations are martingales, so the fair price
should change as a martingale.

\section{Deriving the fair price PDE}
 
It is useful to think of the conditional estimate of the fair
price at some point during a weather contract as being dependent on the conditional estimate of the expected
index, rather than, say, on the temperature so far during the contract.
The former is much more convenient in many situations
because the expected index is a Brownian motion while the temperature is very highly
autocorrelated and hard to model. Thus we have:

\begin{equation}
  V=V(\mu,\sigma_x)
\end{equation}
where $\mu$ and $\sigma_x$ are the conditional mean and standard deviation 
of the settlement index.

Differentiating this with respect to time gives:
\begin{eqnarray}\label{dv}
  dV&=&\frac{\partial V}{\partial \mu}d\mu
     +\frac{\partial V}{\partial \sigma}d\sigma
     +\frac{1}{2}\frac{\partial^2 V}{\partial \mu^2}d\mu^2+...\\\nonumber
     &=&\delta d\mu
     +\zeta d\sigma
     +\frac{1}{2}\gamma d\mu^2+...\\\nonumber
     &=&\delta \sigma dB
          +\theta dt
     +\frac{1}{2}\gamma \sigma^2 dB^2+...\\\nonumber
     &=&\delta \sigma dB
     +\theta dt
     +\frac{1}{2}\gamma \sigma^2 dt+...\\\nonumber
     &= &\delta \sigma dB
     +dt (\theta+\frac{1}{2}\sigma^2\gamma)+...\\\nonumber
     &\approx &\delta \sigma dB
     +dt (\theta+\frac{1}{2}\sigma^2\gamma)
\end{eqnarray}

This last expression is the Ito derivative, and we have used the standard definitions
$\delta=\frac{\partial V}{\partial \mu}$,
$\zeta=\frac{\partial V}{\partial \sigma}$,
$\theta=\frac{\partial V}{\partial t}$
and
$\gamma=\frac{\partial^2 V}{\partial \mu^2}$ (see~\citet{jewson03b}).
We see that changes in $V$ are driven by stochastic jumps (the $dB$ term) 
and by a deterministic drift (the $dt$ term).

But we have already seen in section~\ref{fpismart} that $V$ is a martingale, and hence that there can be no drift
term. Thus we must have that the coefficient of $dt$ in this equation is zero, giving:

\begin{equation}
  \theta+\frac{1}{2}\sigma^2\gamma=0
\end{equation}

or, re-expanding in terms of the full notation:
\begin{equation}\label{pde}
  \frac{\partial V}{\partial t}+\frac{1}{2}\sigma^2\frac{\partial^2 V}{\partial \mu^2}=0
\end{equation}

We conclude that the fair price of a weather option satisfies a PDE, which is a backwards diffusion-type of equation.
The diffusion coefficient comes from the volatility of weather forecasts.

\subsection{Relation to Feynmann-Kac}

There is a close relation to what we have just derived, and the Feynmann-Kac theory discussed
in section~\ref{fk}. Equation~\ref{pde} is a particular example of equation~\ref{fkpde}.
Applying the Feynmann-Kac theorem, we see that we can solve this equation by
integrating a stochastic process and taking an expectation. The stochastic process is given
by equation~\ref{bm}. We have come full circle.

\section{The stochastic process for the fair price and the VaR}

Given the PDE for the fair price (equation~\ref{pde}) equation~\ref{dv} now simplifies to:
\begin{equation}\label{dv2}
  dV=\delta \sigma dB
\end{equation}

In other words, changes in the fair price of a weather derivative over short time horizons are normally 
distributed around the current fair price, and have a volatility given simply
in terms of the $\delta$ of the contract, and the $\sigma$ of the underlying expected index.

Holders of financial contracts often like to know how much they could
lose, and how quickly. Minimising, or limiting, the amount one could lose
is known as risk management. One of the most common measures of risk
used in risk management is the market value at risk, or market VaR. 
Market VaR is defined to be the 5\% level of the distribution of possible market
prices of a contract at a specified future time. 
So far we haven't considered market prices, just fair value, so we will
define \emph{actuarial VaR} as the 5\% level of the distribution of
possible \emph{fair} values of a contract at a specified future time.

Equation~\ref{dv2} actually gives us the actuarial VaR over short time horizons
\footnote{this is the useful result we mention in the abstract}.
The distribution of changes in the fair value is given by:

\begin{equation}\label{dv3}
  dV \sim N(0,\delta \sigma)
\end{equation}

and the VaR is a quantile from this distribution.

This equation doesn't apply over finite time horizons because both $\delta$ and
$\sigma$ change with time. We could integrate equation~\ref{dv2} forward in time
to get around this and derive the distribution of $dV$ over finite time horizons.
However, very small time steps are needed because $\delta$ can change very rapidly in 
certain situations.
It is generally easier to calculate actuarial VaR over longer horizons by integrating
the underlying process (equation~\ref{bm}) to give a distribution of future values of the expected
index (which can be done analytically) and \emph{then} calculating the distribution
of values of $V$. These issues are discussed in more details in~\citet{jewson03k}.

\section{Arbitrage pricing, Black-Scholes and Black}

Equity options are contracts that have a payoff dependent on the future 
level of some share price. By trading the shares themselves very frequently
one can more or less replicate the payoff structure of an equity option.
The cost of doing this replication tell us what the value of the option should be.
This is the Black-Scholes theory of option pricing~\citep{blacks73}.
If the replication is done using forward contracts on the equity rather than 
the equity itself then the Black-Scholes theory must be modified as described
by~\cite{black76} (although if we set interest rates to zero, as we will, the two theories
are the same).

We can apply the Black-Scholes argument to weather options, as described in~\citet{jewsonzervos03a}.
\footnote{thanks to Anna-Maria Velioti for pointing out that equation 28 in this paper
has a sign error in the last term}
We imagine hedging the weather option using weather forwards. The final equation is
slightly different from the Black equation, because the underlying process we use to model
the expectation of the weather index is slightly different from the processes that are
used to model share prices. In the case where interest rates are zero we find that 
the Black equation
for weather is the same as the PDE for the fair price, already given in equation~\ref{pde}.

To summarise: in an actuarial pricing world equation~\ref{pde} gives us the fair price for weather options.
In a Black-Scholes world (where frequent trading is possible) equation~\ref{pde} gives
us \emph{both} the fair price \emph{and} the market price.

\section{Notes on the relation between Black-Scholes and Weather Black-Scholes}

In the standard Black-Scholes world one way of expressing the price of an option is as the
discounted expected payoff with expectations calculated \emph{under the risk neutral measure}.
A curious feature of the weather derivative version of the Black-Scholes model is that the change
of measure is not necessary, since the underlying process (given by equation~\ref{bm}) does not
have any drift. The objective measure is the risk neutral measure. 

\section{Summary}

We have argued that temperature forecasts for a fixed target day
can be modelled as Brownian motion with deterministic volatility.
We have also argued that temperature forecasts for a fixed target month
can be modelled as a Brownian motion, but with a more complex volatility structure.
Standard results for Brownian motion thus apply such as the forward and backward
Fokker-Planck equations, and equations for transition probabilities.

We have also argued that the fair price of a weather derivative must be a martingale.
As a result, the fair price satisfies both a PDE and an SDE, and the SDE
can be used to estimate actuarial VaR over short time horizons.

Finally, under additional assumptions about the weather market the PDE for the fair price 
is the weather derivative equivalent of the Black-Scholes and Black equations. 

\bibliography{feynmannkac}

\section{Legal statement}

The author was employed by RMS at the time that this article was written.

However, neither the research behind this article nor the writing of this
article were in the course of his employment,
(where 'in the course of his employment' is within the meaning of the Copyright, Designs and Patents Act 1988, Section 11),
nor were they in the course of his normal duties, or in the course of
duties falling outside his normal duties but specifically assigned to him
(where 'in the course of his normal duties' and 'in the course of duties
falling outside his normal duties' are within the meanings of the Patents Act 1977, Section 39).
Furthermore the article does not contain any proprietary information or
trade secrets of RMS.
As a result, the author is the owner of all the intellectual
property rights (including, but not limited to, copyright, moral rights,
design rights and rights to inventions) associated with and arising from
this article. The author reserves all these rights.
No-one may reproduce, store or transmit, in any form or by any
means, any part of this article without the author's prior written permission.
The moral rights of the author have been asserted.

\end{document}